%% file: main.tex
\documentclass[runningheads]{llncs}
\usepackage{graphicx}
\usepackage{amsmath,verbatim,amssymb,amsfonts,amscd,graphicx}
\usepackage{hyperref}
\usepackage[english]{babel}
\usepackage[utf8]{inputenc}
\usepackage{tikz}
\usepackage{amsmath}
\usepackage{mathtools}
\usepackage{graphicx}
\usepackage{subcaption}
\usepackage{dcolumn}
\usepackage{makecell}
\usepackage{array}
\newcolumntype{d}[1]{D{.}{.}{#1}}
\usetikzlibrary{shapes,positioning,arrows.meta}
\usetikzlibrary{shapes.geometric,calc,matrix,arrows,decorations.pathmorphing}
\usetikzlibrary{3d,shapes,snakes}
\usetikzlibrary{shadings}
\tikzset{
	braces/.style = {
		outer sep=-1pt,
		left delimiter=(,
		right delimiter=),
		align=center,
	},
}

\usepackage{pgfplots}
\pgfplotsset{compat=newest}

\DeclareMathOperator{\R}{\mathbb{R}}

\newcommand{\cTT}{T} 
\newcommand{\cYY}{Y} 
\newcommand{\cT}{\mathcal{T}} 

\newcommand{\SQN}{\ensuremath{\mathrm{S}q\mathrm{N}}}

\newcommand{\NGF}{\ensuremath{\mathrm{NGF}}}
\newcommand{\SSD}{\ensuremath{\mathrm{SSD}}}

\newcommand{\ie}{{i.e.}}

\newcommand{\argmax}{\mathop{\mathrm{arg \ max}}}

\newcommand{\Sprod}[2][]{\left\langle#2\right\rangle_{#1}}
\newcommand{\rank}{\mathop{\mathrm{rank}}}

\usepackage{xcolor}
\usepackage{todonotes}
\newdimen\fwd

\newsavebox{\JMparbox}
{\end{minipage}\end{lrbox} \box\JMparbox\par\vskip1mm}

\begin{document}

\title{Variational registration of multiple images with the SVD based \SQN\ distance measure}

\titlerunning{Variational registration with S$q$N}
%
\author{Kai Brehmer\inst{1} \and
Hari Om Aggrawal\inst{1} \and
Stefan Heldmann\inst{2} \and
Jan Modersitzki\inst{1,2}
}

\authorrunning{K. Brehmer et al.}
%
\institute{Institute of Mathematics and Image Computing, University of Lübeck, Germany 
\email{brehmer@mic.uni-luebeck.de} 
\and
Fraunhofer Institute for Digital Medicine MEVIS, Lübeck, Germany}
%
\maketitle              
%

\begin{abstract}
Image registration, especially the quantification of image similarity, is an important task in image processing. Various approaches for the comparison of two images are discussed in the literature. However, although most of these approaches perform very well in a two image scenario, an extension to a multiple images scenario deserves attention.
In this article, we discuss and compare registration methods for multiple images. Our key assumption is, that information about the singular values of a feature matrix of images can be used for alignment. We introduce, discuss and relate three recent approaches from the literature: the Schatten $q$-norm based \SQN\ distance measure, a rank based approach, and a feature volume based approach. 
We also present results for typical applications such as dynamic image sequences or stacks of histological sections. Our results indicate that the \SQN\ approach is in fact a suitable distance measure for image registration. Moreover, our examples also indicate that the results obtained by \SQN\ are superior to those obtained by its competitors.

\keywords{Groupwise registration, Dynamic Imaging, 3D reconstruction}
\end{abstract}

\input{01-Introduction.tex} 
\input{02-Methods.tex}

\input{03-Numerics.tex}

\input{04-Results.tex}

\input{05-Discussion.tex}

%
%

\bibliographystyle{splncs04}
\bibliography{2019-SSVM}

\end{document}

%% file: 01-Introduction.tex

\begin{section}{Introduction}

Typical applications in medical imaging are to analyze spatio-temporal variations of bio-medical images. A prerequisite for such analysis is that images are aligned and in many cases joint registration of multiple images is required. Examples are, e.g., analysis of images from different time points and/or different complimentary modalities, atlas registration, longitudinal normalization, motion correction or image reconstruction \cite{BhatiaEtAl2004,CootesEtAl2004,Geng2009,Guyader2018,HuizingaEtAl2016,Joshi2004,PolflietEtAl2018,Schmitt2006,YigitsoyWachingerNavab2011}

A number of registration models are already available to register a pair of two images~\cite{Modersitzki2009,SotirasDavatzikosParagios2013,ZitFlus2003}, but their simple extension to register a group of images might suffer from various problems. Generally, these pair-wise methods assume one of the images as a reference image, and therefore registrations are implicitly biased towards the reference image. Moreover, the selection of a reference image from the given image sequence is not always a very straight forward process. Most importantly, these registration models are primarily influenced by features shared by the image pair and less affected by the features other images have in the image sequence. Therefore, this approach does not account the global information available in the image sequence. It has also been shown that these methods have slow convergence rate compared to the groupwise methods~\cite{Brehmer2018novel,Brehmer2018}.
 
To avoid the selection of a reference image and the related bias, Joshi et al.~\cite{Joshi2004} proposed the registration of each image from the image sequence with respect to the group mean of the registered image sequence. This approach does not need to define the reference image explicitly, moreover accounts the global information through the group mean. This approach inherits the assumption that every image in the image sequence is almost similar to the group mean. 

Recently, Guyader~\cite{Guyader2018} and Brehmer \cite{Brehmer2018novel,Brehmer2018} proposed groupwise registration methods for a sequence of images. The underlying assumption is that images are linearly dependent if they are aligned. The linear dependency idea completely circumvents the need of defining a group mean image. Both of these methods construct an image matrix where each column is corresponding to an image from the sequence. Brehmer~\cite{Brehmer2018novel,Brehmer2018} estimates transformation fields by minimizing the rank of the matrix and implicitly forcing columns of the matrix to become linear dependent to each other. 
Guyader~\cite{Guyader2018} utilizes the multivariate version of mutual information, called total correlation, to define a groupwise registration model. 

The paper is structured as follows: In Section \ref{regmodels}, we discuss mathematical formulations of SVD based image registration approaches. More precise, we discuss a general framework for groupwise registration models based on correlation maximization. In Section \ref{sec:numerics} we briefly discuss the used numerical setting. After that, in Section \ref{results}, we demonstrate the performance of some of the proposed methods on two datasets and compare them with other state-of-the-art methods.



\end{section}

%% file: 02-Methods.tex
\begin{section}%
	{Registration approaches for multiple images}%
	\label{regmodels}






In this section, we describe our Schatten $q$-norm based distance measure $\SQN$ for multiple images. We start by briefly outlining a standard variational registration framework for two images~\cite{Modersitzki2009}. We then present a straightforward extension for multiple images and discuss the drawbacks of the naive approach drawbacks. The main drawbacks are its sequential and thus ordering dependent assessment of the image frames and the weak coupling of image information over the frames.

We then present the setting of the \SQN\ distance measure. The main idea is to make use of the singular values of an image feature array. Finally, we relate the Schatten $q$-norm based distance measure to work of Friedman et al.~\cite{Friedman1981} and Guyader et. al.~\cite{Guyader2018}.

\begin{subsection}{Variational registration approach for two images}%
	\label{sec:IR-2}

We start the discussion with a standard approach to image registration; see e.g.~\cite{Modersitzki2009} for details. To simplify discussion, an image $\cT$ is assumed to be a real valued intensity function $\cT :\R^d\to\R$ with compact support in a domain $\Omega\subset\R^d$.
Given two images $\cT_0,\cT_1$, the goal of image registration is to find a transformation $y:\R^d\to\R^d$ such that ideally $\cT_1\circ y\approx \cT_0$, where $\cT\circ y(x):=\cT(y(x))$. To achieve this goal, we choose a variational framework where a joined functional 
\begin{equation}
	J^{\mathrm{two}}(y;\cT_0,\cT_1):=D(\cT_0, \cT_1\circ y)+S(y),
\end{equation}
is to be minimized over an admissible set of transformations. Various choices for distance measures $D$ and regularizers $S$ are discussed in the literature; see e.g. \cite{Modersitzki2009} and references therein. A thorough discussion is beyond the scope of this paper. Here, we only briefly recall the $L_2$-norm (sum of squared distances, \SSD), the normalized gradient field (\NGF)~\cite{Haber2006}, and the elastic potential~\cite{FisEls1973}:
\begin{eqnarray}
    \label{eq:DM-SSD}
    D^{\mathrm{SSD}}(\cT_0, \cT_1\circ y)&:=& \textstyle 
	\frac12 \|\cT_1\circ y-\cT_0\|_{L_2(\Omega)}^2,
    \\
    \label{eq:DM-NGF}
    D^{\mathrm{NGF}}(\cT_0, \cT_1\circ y)
    &:=& \textstyle 
    \frac12\int_{\Omega} \Big[1 -\Sprod{ 
		\frac{\nabla\cT_1\circ y}{\|\nabla\cT_1\circ y\|_{\eta}}, 
    	\frac{\nabla\cT_0 }{\|\nabla\cT_0\|_{\eta}}}^2\Big]\ dx
    \\  
    \label{eq:REG-ELAS}
    S^{\mathrm{elas}}(y) &:=& 
    \textstyle 
    \frac12\|  \mu \, \mathrm{tr}(E^2) + \lambda \,\mathrm{tr}(E)^2 \|_{L_2(\Omega)}^2 
\end{eqnarray}
with $\|a\|_\eta:=\sqrt{\Sprod{a,a}+\eta}$, $\eta>0$ and strain $E := \nabla y + \nabla y^\top - I$ where $I$ is the identity matrix.

Derivations of image intensities are also commonly used to quantify image similarity. For a unified conceptual framework, we introduce a feature map $F$ that maps an image to a Hilbert space of features.
Any metrics $\mu$ on the feature space can then be used for registration:
$D(\cT_0,\cT_1):=\mu(F(\cT_0),F(\cT_1))$. Examples of such feature maps are e.g. intensity normalization 
$F^\text{IN}(\cT) = \cT/\|\cT\|_{L_2}$ or the normalized gradient field, 
$F^\text{NGF}(\cT) = \nabla\cT/\|\nabla\cT\|_{\eta}$, to name a few.
Note that the NGF distance measure is based on $\nabla\cT(x)/\|\nabla\cT(x)\|_{\eta}$ whereas the feature map is based on $\nabla\cT(x)/\|\nabla\cT\|_{L_2}$.
\end{subsection}

\begin{subsection}{Sequential registration approach for multiple images}%
	\label{sec:IR-K}

Our goal is to extend the standard registration to sequences of images
$\cTT=(\cT_1,\ldots,\cT_K)$. Note that the images might be given as a time series such as our DCE-MRI example, a structured process such as the HISTO application, or even an unstructured ensemble of images such as an atlas generation.


The first approach is to simply apply the above framework sequentially. With transformations $\cYY=(y_1,\ldots,y_K)$ the corresponding energy to be minimized with respect to $\cYY$ reads
\begin{equation}\label{eq:J-seq}
	J^{\mathrm{seq}}(\cYY;\cTT):=\sum_{k=2}^K\left\{ 
		D(\cT_{k-1}\circ y_{k-1},\cT_{k}\circ y_{k})+S(y_k)\right\}.
\end{equation}
Note that typically, one of the deformations is fixed, e.g., $y_1(x) := x$ for well-posedness.
However, as the problem is usually too big to be solved straightforwardly, a non-linear Gauss-Seidel type iteration is usually applied. Here, one assumes that $\cYY$ is a good starting guess and sequentially improves component by component for $\ell=1,\ldots,K$ by determining optimizers 
\begin{equation}\label{eq:}
z^* \in \arg\min_z
	J^{\mathrm{seq}}(y_1,\ldots,y_{\ell-1},z,y_{\ell+1},\ldots,y_K;\cTT),
\end{equation}
setting $y_\ell:=z^*$ 
and iterates until convergence. This process is generally rather expensive and therefore slow. 
A problem is that the coupling of the different components of $\cYY$ is weak. An update of $y_\ell$ has impact only every $K$-th step in the procedure. Therefore, potentially a high number of iterations is required.
\end{subsection}

\begin{subsection}{Global registration approach for multiple images}%
	\label{sec:IR-G}
	
Here, we propose a registration approach that provides a full coupling of all image frames. Our objective is to find a minimizer $\cYY$ of the energy $J^{\mathrm{glo}}$,
\begin{equation}\label{eq:J-glo}
	J^{\mathrm{glo}}(\cYY;T)
	:=D^{\mathrm{glo}}(\cTT\circ \cYY)+S^{\mathrm{glo}}(\cYY),
\end{equation}
where we use the suggestive abbreviation 
$\cTT\circ \cYY:=(\cT_0\circ y_0,\ldots,\cT_K\circ y_K)$ and for sake of simplicity let be
$
	S^{\mathrm{glo}}(\cYY):=\sum_{k=1}^K S(y_k)
$ 
with $S$ any of the regularizers discussed in Sec.~\ref{sec:IR-2}. Clearly,  
one could debate for a more general or even stronger regularization of $\cYY$. However, this is not in the scope of the paper and we leave the discussion for future work. 
%
The essential contribution is thus the global distance measure that is based on the feature array $F(\cTT):=[F(\cT_1),\ ,\ldots,\ F(\cT_K)]$ which comprises the features of the image sequence and its symmetric, positive semi-definite correlation matrix $C=\Sprod{F,F} \in \R^{K\times K}$ where $C_{ij}$ assembles the correlations of $F(\cT_i)$ and $F(\cT_j)$. 
Note that we assumed $F$ maps into a Hilbert space such that the correlation is well defined according to the corresponding inner product.
%
%
Our key assumption is that the rank of the feature array is minimal if the image frames are aligned. Note that we actually aim to exclude the trivial situation $\rank F=0$ as this implies that all features are zero.
We also note that the assumption may not hold for multi-modal images, if the feature map does not compensate intensity variations. 
Therefore, a plain image intensity based feature map may not be successful. If we expect that intensity changes will occur at similar positions in space, e.g., the NGF feature map is a valid choice.

\end{subsection}

\begin{subsection}{Schatten $q$-norm based image similarity measure $D_{S,q}$}%
	\label{sec:}
	
The above considerations suggest to choose $\rank F$ as a distance measure.
In~\cite{Brehmer2018novel,Brehmer2018}, Brehmer et al. proposed to reformulate the rank minimization problem in terms of a relaxation of the rank function based on a so-called Schatten $q$-norm. Roughly speaking, the Schatten $q$-norm of an operator is the $q$-norm of the vector of its singular values. Thus
\begin{equation}
	D_{S,q}(\cTT):=
	\|F(\cTT)\|_{S,q}:=
	\Big( \sum_{k=1}^{K} \sigma_k(F(\cTT))^q \Big)^{1/q}	
\end{equation}
where $\sigma_k$, $k=1,\ldots,K$, denote the non-zero singular values of $F(\cTT)$. 
Before we discuss numerical details, we relate this measure to other rank based similarity measures for image stacks. Particularly we address volume minimization of the feature parallelotope and correlation maximization of normalized features.
\end{subsection}

\begin{subsection}{Volume minimization of the feature parallelotope}%
	\label{sec:}
	
The above approach can be linked to work of Guyader et. al.~\cite{Guyader2018}.
To this end, we consider the minimization of the volume of the parallelotope spanned by the columns of $F(\cTT)$. Equivalently, we can consider the determinant of $C$ or, exploring the monotonicity of the logarithm, set
\begin{equation}\textstyle
	D(\cTT)
	:=\log(\det(C(\cTT)))
	=\log(\prod_{k=1}^K\sigma_k^2(F(\cTT)))
	=2\sum_{k=1}^K\log(\sigma_k(F(\cTT))).
\end{equation}
This expression is related to the volume of a normalized covariance matrix which is the total correlation in~\cite{Guyader2018} and used
as a similarity measure for group-wise registration.

However, a volume based approach has a severe drawback; see also the discussion in~\cite{HaberModersitzki2006}. To illustrate this, we consider two feature vectors $f_1\ne0$ and $f_2$ with angle $\alpha$. Hence, $\mathrm{volume}(f_1,f_2)=\|f_1\|\|f_2\|\sin\alpha$. This value is minimal if the vectors are linearly dependent. Unfortunately, this also happens if $f_2=0$. In a registration context, this implies that a translation of one of the images, say, about the diameter of $\Omega$ yields a global optimizer. In~\cite{HaberModersitzki2006} it is therefore suggested to replace the minimization of volume by a maximization of correlation $|\cos\alpha|$.
This value is maximal iff and only iff $f_2=\pm f_1$ and is in fact minimal if $f_2=0$. This subtle difference is very important in a registration context.

\end{subsection}

\begin{subsection}{Correlation maximization of normalized features}%
	\label{sec:}
	
In this section we focus on correlation maximization and do not discuss the corresponding minimization formulation. We also assume that feature vectors are normalized, i.e. $\|F(\cT_k)\|=1$. For the correlation matrix $C(\cTT) \in\R^{K,K}$ holds
\begin{equation}
    C_{kk}=1,\quad
	C_{jk}=\Sprod{F(\cT_j),F(\cT_k)}=\cos \gamma_{jk},
\end{equation}
where $\gamma_{jk}$ denotes the angle between the $j$-th and $k$-th feature.
In the two image setting it is therefore natural to maximize $|C_{1,2}|$ if we account both, for positive and negative correlation. This is the underlying idea of normalized cross correlation. Note that the NGF approach is still different as the correlation is computed point wise and finally averaged.

For the multiple image setting, the best scenario is $C\in\{\pm1\}^{K,K}$. If only non-negative correlation is considered, the ideal case is $C(\cTT)=1\cdot1^{\top}$. On the opposite, the worst case scenario for registration is that $C(\cTT)=I$ meaning all features are fully uncorrelated. Therefore, a suitable distance measure is to maximize the difference 
\begin{equation}\label{eq:D-max}
	\textstyle
	D(\cTT)
	:=\|C(\cTT)-I\|_{M},
\end{equation}
where $\|\cdot\|_{M}$ denotes a suitable matrix norm.

\end{subsection}

\begin{subsection}{Correlation maximization and Schatten $q$-norms}%
	\label{sec:}
	
Specifically, choosing $\|\cdot\|_M = \|\cdot\|_{S,q}$ a Schatten $q$-norm in~\eqref{eq:D-max} 
we obtain
\begin{align}\label{eq:SqN-1}\textstyle
	D(\cTT)=\|C(\cTT)-I\|_{S,q}
    =\left(\sum_{k=1}^{K} ( \sigma_k^2(F(\cTT))- 1)^q\right)^{1/q}.
\end{align}
We investigate the special cases $q=2$ and $q=\infty$. Note that 
\begin{eqnarray*}
 	\|A\|_{S,\infty}
	&=&\sigma_{\max}(A),\quad 
	\mbox{the largest singular value of $A$, and}\quad
	\\
 	\|A\|_{S,2}^2
	&=&\sum_k \sigma_k^2
	 =\mathrm{trace}(A^{\top}A)
	 =\sum_{j,k}|a_{j,k}|^2
	 =\|A\|_{\mathrm{Fro}}^2.
\end{eqnarray*}
Thus, choosing the Schatten $\infty$-norm yields maximizing $\sigma^2_{\max}(F(\cTT))-1$. This is equivalent to maximizing  the largest singular value of $F(\cTT)$, see also~\cite{Friedman1981}:
\[
	\argmax \, \|C(\cTT)-I\|_{S,\infty} = \argmax \, \sigma_{\max}(F(\cTT)).
\]
For the Schatten $2$-norm we have $D(\cTT)=\|C(\cTT)-I\|_{S,2}^2=\sum_{i\ne j}|C_{ij}|^2$ 
which shows that the distance is quadratic mean of the correlation among the image features.
Furthermore, a direct computation shows 
\begin{equation*}
	D(\cTT) = \|C(\cTT)-I\|_{S,2}^2 = \|F\|_{S,4}^4 - K.
\end{equation*}
Here, we exploit the special structure of correlation matrix $C$, \ie, $\mathrm{trace}(C) = K$.


To this end, we define the two \SQN\ distance measures for NGF features as follows:
\begin{eqnarray}
    \SQN_4(T) &:=& K - \|F^\NGF(T)\|_{S,4}^4
    \\
    \SQN_\infty(T) &:=& -\sigma_{\max}(F^{\NGF}(T))
\end{eqnarray}

\end{subsection}

%
%


\end{section}

%% file: 03-Numerics.tex
\begin{section}{Numerical methods}\label{sec:numerics}

For the optimization of the functional $J^{\mathrm{seq}}$ (cf.~\eqref{eq:J-seq}) we use the discretize-then-optimize framework introduced in~\cite{2006-SISC-HM}. The basic concept is to use a sequence of discretized finite dimensional optimization problems. A smooth approximation of the problem is represented with few degrees of freedom. It is expected that the optimization is fast as the problem is low dimensional and smooth. Its numerical solution is prolongated and then serves as a starting guess for the finer resolved problem. It is expected that a numerical solution can be computed fast, as the starting point is expected to be close to the solution. The process is generally terminated when reaching the resolution of the given data. Note that the images are only smoothed in the spatial domain.

To solve the discrete problem on a fixed resolution we use a quasi-Newton type approach. More precisely, we use L-BFGS with the Hessian of the regularizer as an initial approximation of the metric and a Wolfe linesearch; see, e.g.~\cite{NocedalWright2006} for optimization and \cite{Modersitzki2009} for details.

For the optimization of $J^{\SQN}$,
\begin{equation}\label{eq:J-SqN}
	J^{\SQN}(\cYY;T)
	:=\SQN(\cTT\circ \cYY)+S^{\mathrm{glo}}(\cYY),
\end{equation}
we use similar concepts as above for the regularization term.

For the \SQN\ distance, we remark that the distance is a rather simple algebraic expression of the singular values of the feature matrix. The challenging part is thus the derivative of the singular values. Here, we follow~\cite{Papadopoulo2000}. A singular value decomposition of the feature matrix $F \in \R^{n\times K}$ is denoted by $F=U\Sigma V^{\top}$, where the matrices $U=(u_{i,k})\in\R^{n,n}$ and $V=(v_{j,k})\in\R^{K,K}$ are orthogonal and $\Sigma\in\R^{n,K}$ is a non-negative diagonal matrix with the singular values $\sigma_k(F)$ as diagonal entries.
From~\cite{Papadopoulo2000} we have the surprisingly simple relation\\
$\frac{\partial\sigma_k(F)}{\partial F_{i,j}} = u_{i,k} v_{j,k}$
that is used in our implementation.

\end{section}

%% file: 04-Results.tex
\section{Results}\label{results}
We now present results for the registration of histological serial sectioning of a marmoset monkey brain as well as for DCE-MRI sequences of a human kidney. For the given datasets, we will compare the registration results of $\SQN_4$, $\SQN_\infty$ in comparison to a total correlation based approach like in \cite{Guyader2018} and sequential \NGF.
We start with registrations of a serial sectioning of a marmoset monkey brain; data courtesy of Harald Möller, Max Planck Institute for Human Cognitive and Brain Sciences, Leipzig, Germany \cite{Marschner2014}.
The dataset consists of every 4th slice of the original serial sectioning of the brain, in total 69 slices of sizes from $2252 \times 3957$ pixels up to $7655 \times 9965$ pixels. For proof of concept we reduced the number of pixels per slice to reduce computation time to a reasonable level.
The objective of the registration of histological slices is to align them in order to reconstruct the $3D$ volume of the tissue.

\fwd=17mm
\newbox\myImg
\setbox\myImg=\hbox{\includegraphics[height=\fwd,angle=0]{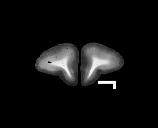}}	
\begin{figure}[t]\centering
	\setlength{\tabcolsep}{10pt}
	\begin{tabular}{ccc}
		Slice 5 & Slice 30 & Slice 46
		\\
		\includegraphics[height=\fwd]{./fig/histo_5}
		&   \includegraphics[height=\fwd]{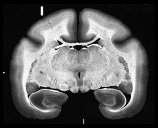}
		&   \includegraphics[height=\fwd]{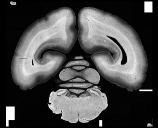}
	\end{tabular}
	
	\caption{%
		Three representative axial slices of a marmoset monkey brain dataset; data courtesy of Harald Möller \cite{Marschner2014}
	}
	\label{fig:overviewMBP}
\end{figure}

Fig.~\ref{fig:overviewMBP} shows three representative axial slices of the data set. The main difficulties of registering this particular dataset are the different sizes of the slices on the one hand and the translation of whole parts of the imagestack within the domain on the other hand. Furthermore we didn't use a pre-segmentation of the dataset to show robustness of the registration approaches against artifacts in the background region. The background region of the slices contains several markings of the examiners like white rectangles as well as dust and dirt from the object slide captured during the high resolution scanning process; see Fig.~\ref{fig:overviewMBP}.

Fig.~\ref{fig:resultsMBP} shows two sagittal slices (top and bottom row) through the image stack from the reduced, unregistered monkey brain dataset besides the registration results to illustrate the alignment of the slices. As expected the results of $\SQN_4$ are quite similar to the results of $\SQN_\infty$.
The computation for the groupwise approaches using $\SQN$ as well as the total correlation approach from \cite{Guyader2018} took about 45 to 50 minutes for a resolution of $128 \times 158$ pixels for each of the 69 slices. Compared to this, the sequential NGF approach with just one sweep needed about 2.2 times the computation time (ca. 110 minutes). However, from visual comparison it is obvious that many more sweeps are needed to achieve results comparable to those of the groupwise approaches; see Fig.~\ref{fig:resultsMBP}. Everything was implemented in Python using Numpy and Scipy for optimization.

Moreover, we used a random permutation of the stack of histological serial sections to demonstrate invariance to the order of images of the singular value based groupwise registration approaches. We randomly permuted the order of images, registered the stack in random order using $\SQN_4$ and reordered it afterwards; see Fig.~\ref{fig:resultsMBP_permute}, center column. As expected, the results are the same as for registration using $\SQN_4$ without random permutation; cf. Fig.~\ref{fig:resultsMBP} and Fig.~\ref{fig:resultsMBP_permute} for comparison.

Next we present registration results for a DCE-MRI sequence of a human kidney; data courtesy of Jarle R{\o}rvik, Haukeland University Hospital Bergen, Norway.
Here, 3D images are taken at 45 time points. For ease of presentation and to have a reasonable level of computation time we show results for a 2D slice over time. More precisely, we use 178-by-95 coronal slices of a 178-by-95-by-30-by-45 volume for z-slice 18; see Fig.~\ref{fig:overviewKidney} for representative slices. All time points are used for registration.
The objective here is to register the slices while maintaining the dynamics.
Fig.~\ref{fig:resultsKidney} illustrates the stack of slices for the different registration approaches using a sagittal cut through the stack, analog to the results for the histological serial sections shown in Fig.~\ref{fig:resultsMBP}.
The illustrated results were achieved using three different levels of spatial resolution up to half the original resolution in about 8 minutes per groupwise approach. The result of the sequential approach was achieved in about twice the time using just one sweep. For the alignment using the approach from \cite{Guyader2018}, we couldn't find a parameter setting to achieve results comparable to the \SQN- approaches.

\fwd=15mm
\newcommand{\hfs}{\scriptsize}
\begin{figure}[t]\centering
	\setlength{\tabcolsep}{10pt}
	\begin{tabular}{c c c c c c}
		& \hfs{Unregistered} & \hfs{$\SQN_4$} & \hfs{$\SQN_\infty$} & \hfs{\makecell{Total \\ Correlation}} & \hfs{NGF}
		\\
		\rotatebox[]{90}{\hskip-25mm \hfs{Position 53}}
		& \includegraphics[height=\fwd,angle=-90]{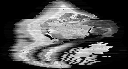}
		&   \includegraphics[height=\fwd,angle=-90]{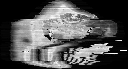}
		&   \includegraphics[height=\fwd,angle=-90]{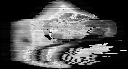}
		&	\includegraphics[height=\fwd,angle=-90]{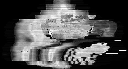}
		&	\includegraphics[height=\fwd,angle=-90]{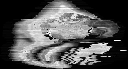}
		\\
		\rotatebox[]{90}{\hskip-25mm \hfs{Position 82}} 
		& \includegraphics[height=\fwd,angle=-90]{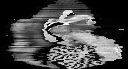}
		&   \includegraphics[height=\fwd,angle=-90]{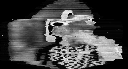}
		&   \includegraphics[height=\fwd,angle=-90]{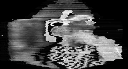}
		&	\includegraphics[height=\fwd,angle=-90]{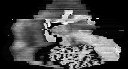}
		& 	\includegraphics[height=\fwd,angle=-90]{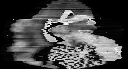}
	\end{tabular}
	\caption{%
		Registration results for 3D reconstruction of the monkey brain datasets. For illustration, we show only 2D slices that are sagittal cuts at two positions, \ie, 53 and 82. 
	}
	
	\label{fig:resultsMBP}
	
\end{figure}

\newbox\myImg
\setbox\myImg=\hbox{\includegraphics[height=\fwd,angle=0]{./fig/histo_image_top53_original}}	
\begin{figure}[t]\centering
	\setlength{\tabcolsep}{10pt}
	\begin{tabular}{cccc}
		& \hfs{Original} & \hfs{Permuted} & \hfs{$\SQN_4$}
		\\
		\rotatebox[origin=c]{90}{\hskip-25mm \hfs{Position 53}}
		&	\includegraphics[height=\fwd,angle=-90]{./fig/histo_image_top53_original}
		&   \includegraphics[height=\ht\myImg,width=\wd\myImg,angle=-90]{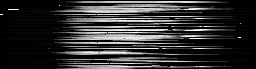}
		&   \includegraphics[height=\fwd,angle=-90]{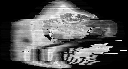}
		\\
		\rotatebox[origin=c]{90}{\hskip-25mm \hfs{Position 82}}
		&	\includegraphics[height=\fwd,angle=-90]{./fig/histo_image_top82_original}
		&   \includegraphics[height=\ht\myImg,width=\wd\myImg,angle=-90]{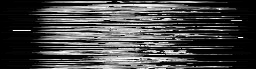}
		&   \includegraphics[height=\fwd,angle=-90]{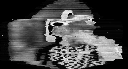}
	\end{tabular}
	
	\caption{%
		Registration results after random permutation of the axial slices. As expected, the results are the same as for the non-permuted image stack; also see Fig.~\ref{fig:resultsMBP} for comparison.
	}
	\label{fig:resultsMBP_permute}
\end{figure}

\fwd=28mm
\newbox\myImg
\setbox\myImg=\hbox{\includegraphics[height=\fwd]{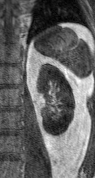}}	
\begin{figure}[t]\centering
	\setlength{\tabcolsep}{10pt}
	\begin{tabular}{ccc}
		\hfs{Time point 5} & \hfs{Time point 11} & \hfs{Time point 21}
		\\
		\includegraphics[height=\fwd]{./fig/kidney_5}
		&   \includegraphics[height=\fwd]{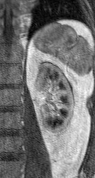}
		&   \includegraphics[height=\fwd]{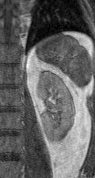}
	\end{tabular}
	
	\caption{%
		Three representative 2D coronal slices of the 4D DCE-MRI dataset of a human kidney; data courtesy of Jarle R{\o}rvik, Haukeland University Hospital, Bergen, Norway.
		The slices are shown at three different time points. The dataset is a 178-by-95-by-30-by-45 volume, the shown slices are 178-by-95.
	}
	\label{fig:overviewKidney}
\end{figure}

\fwd=14mm
\newbox\myImg
\setbox\myImg=\hbox{\includegraphics[height=\fwd,angle=0]{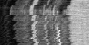}}	
\begin{figure}[t]\centering
	\setlength{\tabcolsep}{10pt}
	\begin{tabular}{cccccc}
		& \hfs{Unregistered} & \hfs{$\SQN_4$} & \hfs{$\SQN_\infty$} & \hfs{\makecell{Total \\ Correlation}} & \hfs{NGF}
		\\
		\rotatebox[origin=c]{90}{\hskip-25mm \hfs{Position 29}}
		& \includegraphics[height=\fwd,angle=-90]{./fig/kidney_image_top29_orig}
		&   \includegraphics[height=\fwd,angle=-90]{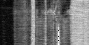}
		&   \includegraphics[height=\fwd,angle=-90]{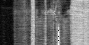}
		&	\includegraphics[height=\fwd,angle=-90]{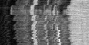}
		&	\includegraphics[height=\fwd,angle=-90]{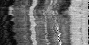}
		\\
		\rotatebox[origin=c]{90}{\hskip-25mm \hfs{Position 40}}
		& \includegraphics[height=\fwd,angle=-90]{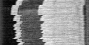}
		&   \includegraphics[height=\fwd,angle=-90]{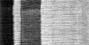}
		&   \includegraphics[height=\fwd,angle=-90]{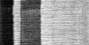}
		&	\includegraphics[height=\fwd,angle=-90]{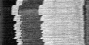}
		& 	\includegraphics[height=\fwd,angle=-90]{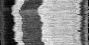}
	\end{tabular}
	
	\caption{%
		Illustrated are sagittal cuts through the stack of 2D slices from a 4D DCE-MRI dataset of a human kidney at positions 29 and 40. The first column shows the unregistered stack. Right next to this the results of the different registration approaches are illustrated. 		
	}
	\label{fig:resultsKidney}
\end{figure}

%% file: 05-Discussion.tex
\section{Discussion and Conclusions}\label{discussion}

The registration of multiple images is an important task in image processing. Conventional approaches often use an extension of a pairwise approach for two images. In this paper, we demonstrate that this approach may come with numerous disadvantages and may be time consuming. We also describe and analyze a recently proposed alternative. The Schatten $q$-norm based \SQN~\cite{Brehmer2018novel,Brehmer2018} distance measure is a reference for our investigations on different singular value based measures such as the maximization of correlation between different images as well as minimization of spanned volumes. For this purpose we have introduced a general formulation using feature maps that map images into Hilbert spaces. This opens a door for even further investigation on image registration methods for multiple images.
With our numerical results we demonstrate that \SQN\ based motion compensation is applicable in dynamic imaging as well as for the alignment of histological serial sections. Moreover, the results clearly show that \SQN\ performs at least as good as standard approaches from the literature. In our experiments both the alignment and the computation time of the groupwise approaches were closer to a desirable solution than the sequential approach using pairwise \NGF.

Furthermore, we outlined that a singular value based approach exploits the global information of a dataset, which cannot be achieved by using two-neighbourhoods in registration. In some specific applications, such as dynamic imaging or reconstruction of histological volumes from serial sections, this can avoid unwanted effects like the so-called banana-effect.
Future work will address the optimal choice of the parameter $q$ and investigations of different variants of feature maps. Finally, different regularization strategies will be investigated.

\subsection*{Acknowledgement}
The authors acknowledge the financial support by the Federal Ministry of Edu\-cation and Research of Germany in the framework of MED4D (project number 05M16FLA)